\documentclass[fleqn,usenatbib]{mnras}
\usepackage{newtxtext,newtxmath}
\usepackage[T1]{fontenc}
\DeclareRobustCommand{\VAN}[3]{#2}
\let\VANthebibliography\thebibliography
\def\thebibliography{\DeclareRobustCommand{\VAN}[3]{##3}\VANthebibliography}

\usepackage{graphicx,multirow}
\renewcommand\apjl{ApJL}

\title[Testing BH area law with inspiral signals]{Hierarchical triple mergers: testing Hawking's area theorem with the inspiral signals}

\author[S.-P. Tang, Y.-Z. Fan and D.-M. Wei]{
Shao-Peng Tang,$^{1}$
Yi-Zhong Fan$^{1,2}$\thanks{E-mail: yzfan@pmo.ac.cn (YZF)}
and Da-Ming Wei$^{1,2}$
\\
$^{1}$Key Laboratory of Dark Matter and Space Astronomy, Purple Mountain Observatory, Chinese Academy of Sciences, Nanjing 210033, China\\
$^{2}$School of Astronomy and Space Science, University of Science and Technology of China, Hefei, Anhui 230026, China
}

\date{Accepted 2023 June 2. Received 2023 May 3; in original form 2023 March 8}

\pubyear{2023}

\begin{document}
\label{firstpage}
\pagerange{\pageref{firstpage}--\pageref{lastpage}}
\maketitle

\begin{abstract}
Hawking's area theorem is one of the fundamental laws of black holes (BHs), which has been tested at a confidence level of $\sim 95\%$ with gravitational wave (GW) observations by analysing the inspiral and ringdown portions of GW signals, independently. In this work, we propose to carry out the test in a new way with the hierarchical triple merger (i.e., two successive BH mergers occurred sequentially within the observation window of GW detectors), for which the properties of the progenitor BHs and the remnant BH of the first coalescence can be inferred from the inspiral portions of the two mergers. As revealed in our simulations, the BH area law can be well confirmed for some plausible hierarchical triple merger events detectable in LIGO/Virgo/KAGRA/LIGO-India's O4/O5 runs.  Our proposed method provides significant facilitation for testing the area law and complements previous methods.
\end{abstract}

\begin{keywords}
black hole physics -- gravitation -- gravitational waves -- methods: data analysis -- black hole mergers.
\end{keywords}

\section{Introduction}
As predicted by \citet{1971PhRvL..26.1344H}, the total horizon area for a system consisting of classical black holes (BHs) never decreases over time, which is known as the black-hole area law (or Hawking's area theorem). The prospects of testing this fundamental theorem with gravitational wave (GW) data have been extensively discussed in the literature \citep{2005ApJ...623..689H, 2016JCAP...10..001G, 2018PhRvD..97l4069C}. The detection of the GW signal from GW150914, the merger of a binary stellar-mass black hole (BBH), opens the era of GW astronomy \citep{2016PhRvL.116f1102A} and has been adopted to directly test the BH area law \citep{2021PhRvL.127a1103I, 2022PhRvD.105f4042K}. The basic idea for such a test is that the inspiral and ringdown portions of some GW signals can independently give the source properties of the progenitor BHs and the remnant BH; therefore, it is possible to measure the change in the total horizon area caused by the merger.

\begin{figure}
\centering
\includegraphics[width=0.48\textwidth]{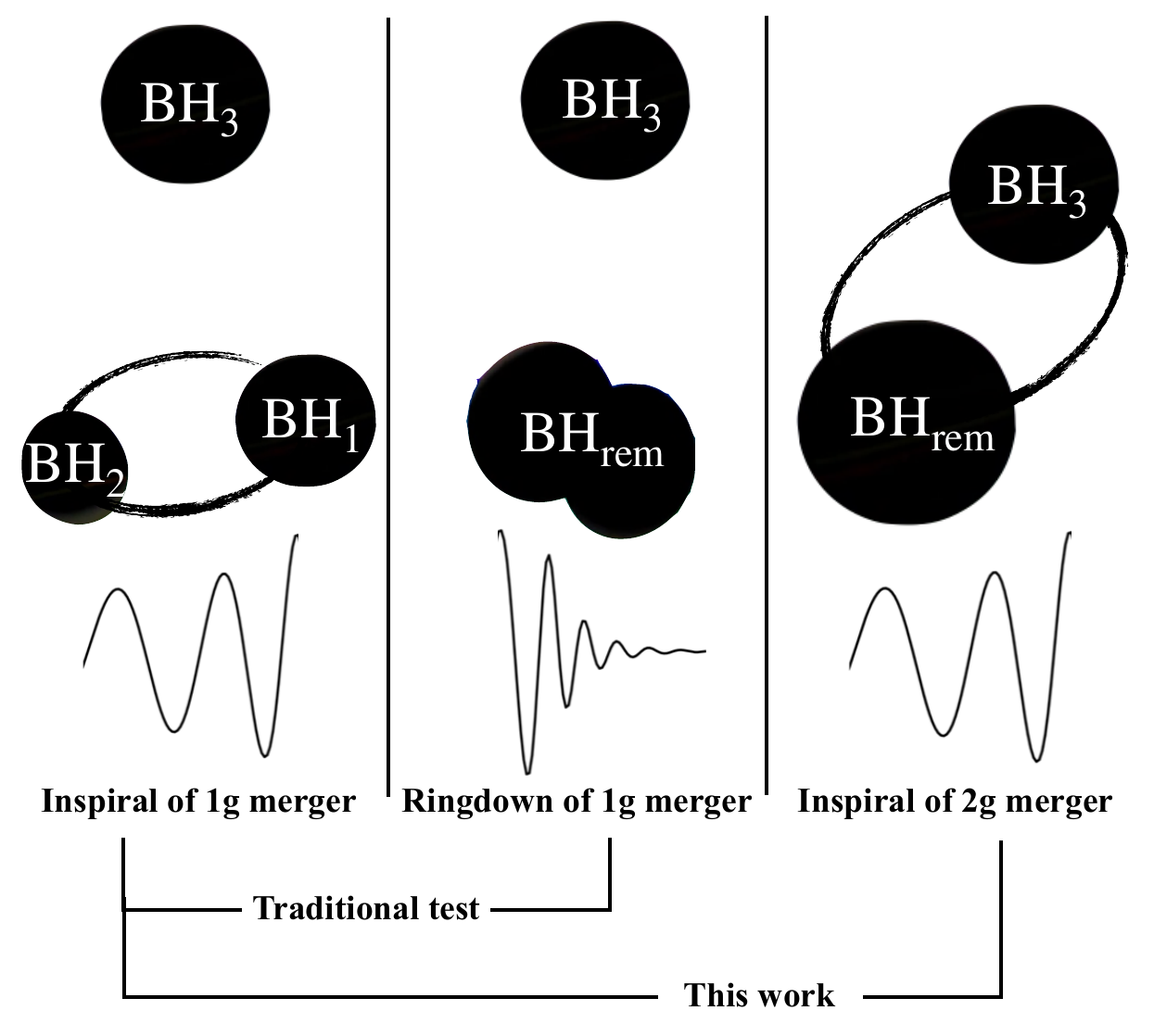}
\caption{Illustration of the two ways to test the BH area law.}
\label{fig:hierarchical-triple-merger}
\end{figure}

\begin{table*}
\caption{Injection configurations and prior distributions}
\label{tb:metadata}
\begin{tabular}{lcccc}
\hline
\multirow{2}{*}{Parameter names}                                  & \multirow{2}{*}{Symbols}    & \multicolumn{2}{c}{Injection configurations}    & \multirow{2}{*}{Prior distributions} \\ \cline{3-4}
                                                                  &                             & 1g merger         & 2g merger                   & \\ \hline
Primary mass                                                      & $m_1\,(M_\odot)$            & 30                & 47.66                       & $5\leq m_1 \leq 100$ \\
Secondary mass                                                    & $m_2\,(M_\odot)$            & 20                & 35                          & $5\leq m_2 \leq 100$ \\
Chirp mass                                                        & $\mathcal{M}\,(M_\odot)$    & --                & --                          & \multirow{2}{*}{UniformInComponents$^a$}  \\
Mass ratio                                                        & $q$                         & --                & --                          & \\
Luminosity distance                                               & $d_{\rm L}\,(\rm Mpc)$      & 1000              & 1000                        & UniformSourceFrame$^b$ \\
Geocentric time                                                   & $t_{\rm c}\,(\rm s)$        & $1\,126\,259\,642.413$    & $1\,189\,331\,642.413$              & U($t_{\rm c}^{*}-0.1$, $t_{\rm c}^{*}+0.1$)$^c$ \\
Right ascension (RA)                                            & $\alpha$                    & 1.375             & 1.375                       & U(0, $2\pi$)/Fixed \\
Declination (Dec.)                                               & $\delta$                    & -1.2108           & -1.2108                     & Cosine/Fixed \\
Coalescence phase                                                 & $\phi$                      & 1.3               & 1.8                         & U(0, $2\pi$) \\
Polarization angle                                                & $\psi$                      & 2.659             & 1.659                       & U(0, $\pi$) \\
Dimensionless spin magnitude of BH$_{1}$                          & $a_1$                       & 0.1               & 0.687                       & U(0, 0.89) \\
Dimensionless spin magnitude of BH$_{2}$                          & $a_2$                       & 0.05              & 0.1                         & U(0, 0.89) \\
Tilt angle between BH$_{1}$'s spin and $\boldsymbol{L}$$^d$       & $\theta_1$                  & 0.5               & 0.1                         & Sine \\
Tilt angle between BH$_{2}$'s spin and $\boldsymbol{L}$           & $\theta_2$                  & 1.0               & 0.8                         & Sine \\
Azimuthal angle separating the BHs' spins                         & $\phi_{12}$                 & 1.7               & 2.0                         & U(0, $2\pi$) \\
Azimuthal position of $\boldsymbol{L}$ on its cone about $\boldsymbol{J}$$^d$ & $\phi_{\rm JL}$ & 0.3               & 2.0                         & U(0, $2\pi$) \\
Inclination between $\boldsymbol{J}$ and the line of sight        & $\theta_{\rm JN}$           & 0.4               & 0.5                         & Sine \\ \hline
\multicolumn{5}{l}{$^a$ The distributions satisfy that the joint distribution of ($m_1$,$m_2$) is uniform (see \href{https://lscsoft.docs.ligo.org/bilby/api/bilby.gw.prior.html\#module-bilby.gw.prior}{the documents} of {\sc Bilby}).}\\
\multicolumn{5}{l}{~~~~The ranges for $\mathcal{M}$ and $q$ are [$\mathcal{M}^{*}-10$, $\mathcal{M}^{*}+10$] and [0.125, 1], where $\mathcal{M}^{*}$ is the injected values.}\\
\multicolumn{5}{l}{$^b$ The corresponding redshift is uniformly distributed in the comoving volume and source frame time.}\\
\multicolumn{5}{l}{$^c$ $t_{\rm c}$ are uniformly distributed within a $200\,\rm ms$ window centred on their true values \citep{2015PhRvD..91d2003V}.}\\
\multicolumn{5}{l}{$^d$ $\boldsymbol{L}$ and $\boldsymbol{J}$ are the orbital angular momentum and the system's total angular momentum, respectively.}
\end{tabular}
\end{table*}
However, \citet{2022PhRvD.105f4042K} pointed out that some systematic errors caused by ringdown modelling may overstate the confidence in confirming the area law. A robust way to test the area law is to estimate the GW parameters purely from the early inspiral and the late ringdown (well described by a spectrum of quasi-normal modes; \citealt{2005ApJ...623..689H}), but the events observed with high signal-to-noise ratios (SNRs) in both stages are rare \citep{2018CQGra..35a4002G}. In this work, we propose a new strategy to test the BH area law with the so-called hierarchical triple merger (as illustrated in Fig.~\ref{fig:hierarchical-triple-merger}), which happens when three BHs are gravitationally bound, causing two successive BH mergers within a time-scale of $\mathcal{O}(\rm yr)$ \citep{2019MNRAS.482...30S}. Some candidates for this kind of merger, though with low confidence, have been suggested in the literature \citep{2020MNRAS.498L..46V, 2021ApJ...907L..48V}. Therefore, if the (ground-based) GW detectors, such as advanced LIGO \citep[aLIGO;][]{2015CQGra..32g4001L} and advanced Virgo \citep[AdV;][]{2015CQGra..32b4001A}, could catch such GW signals, one can independently measure the source properties of the progenitor BHs and the remnant BH of the first coalescence by using the {\it sole} inspiral data (which avoids the systematic errors caused by ringdown modelling as we mentioned above) of the first (1g) and second generation (2g) mergers, respectively. To evaluate the prospect of testing the BH area law with hierarchical triple mergers, we perform Bayesian inference to recover the GW parameters of injected signals. After reconstructing the source properties of the simulated events, we find that the total horizon area of the 1g BHs and the area of the corresponding remnant BH can be well constrained, which allows a test of Hawking's area theorem for some plausible hierarchical triple mergers detectable by aLIGO/AdV/KAGRA/LIGO-India.

\section{Methods}
We follow the `gating and in-painting' method \citep{2021PhRvD.104f3030Z, 2021arXiv210505238C} used in \citet{2022PhRvD.105f4042K} to perform inspiral-only (or pre-truncation) data analysis on the simulated strains $\boldsymbol{s}=\boldsymbol{h}(\boldsymbol{\theta}^{*})+\boldsymbol{n}$, where $\theta^{*}$ are GW parameters with injection values, $\boldsymbol{h}(\boldsymbol{\theta}^{*})$ are injected detector frame waveforms, and $\boldsymbol{n}$ are Gaussian noises coloured with the power-spectral densities (PSDs) as given by the noise curves from: \href{https://dcc.ligo.org/LIGO-T2000012/public}{https://dcc.ligo.org/LIGO-T2000012/public}. Given the GW parameters $\theta$ with values generated by nest sampler, we excise the residual $\boldsymbol{s}-\boldsymbol{h}(\boldsymbol{\theta})$ after the time ($t_{\rm hmeco}$) corresponding to the hybrid minimum energy circular orbit (hybrid MECO; \citealt{2017PhRvD..95f4016C}), then add a component to the gated times such that the contribution of merger/ringdown to the likelihood is zero. Meanwhile, we adopt a fixed sky location in the inspiral-only analysis, as previously done in \citet{2022PhRvD.105f4042K}. The fixed sky location and `{\it gate-start-time}' (i.e., $t_{\rm hmeco}$) are determined by the parameter values corresponding to the maximum likelihood obtained in the complete inspiral-merger-ringdown (IMR) analysis, and both of them do not vary in the inspiral-only analysis. The IMR analysis is performed as the first step using the standard likelihood of GW parameter inference \citep{2019PASP..131b4503B}, which serves only to find the maximum likelihood values for the sky location and to find the time of hybrid MECO. We consider two configurations for the detector network: one consists of aLIGO, AdV, and KAGRA with `O4high', `O4high\_NEW', and `kagra\_10Mpc' sensitivities, respectively; and the other consists of aLIGO/LIGO-India, AdV, and KAGRA with `AplusDesign', `O5high\_NEW', and `kagra\_128Mpc' sensitivities, respectively. The real data recorded by aLIGO/AdV/KAGRA/LIGO-India may suffer from glitches and non-stationary noises, which may produce biased PSD estimation. However, some powerful tools, e.g., BayesLine \citep{2015PhRvD..91h4034L} and BayesWave \citep{2015CQGra..32m5012C}, have been developed to solve this problem \citep{2020CQGra..37e5002A}. Here, we only consider the ideal case by assuming the PSDs can be well estimated and use the same as injections when we perform GW parameter inference. The low-frequency cutoff $f_{\rm low}$ is set to $15\,{\rm Hz}$, which is appropriate for Design/Plus sensitivities \citep{2022ApJ...928...21K}. Besides, the reference frequency is set to $20\,{\rm Hz}$. A phenomenological frequency-domain waveform template {\sc IMRPhenomXPHM} \citep{2021PhRvD.103j4056P} which incorporates multipoles beyond the dominant quadrupole in the precessing frame, is adopted. Assuming that the 1g/2g mergers originate from the identical hierarchical triple system, we let them share the same position (i.e., $\alpha$, $\delta$, and $d_{\rm L}$) in injections and parameter inference (the 1g and 2g parameters are simultaneously sampled in IMR analyses). The true values used in the injections and the priors adopted in the inference are summarized in Table~\ref{tb:metadata}. 
\begin{table*}
\caption{Representative Tests and the Corresponding Results}
\label{tb:tests}
\renewcommand\arraystretch{1.25}
\begin{tabular}{lcccccc}
\hline
\multirow{2}{*}{Different tests$^a$}    & \multirow{2}{*}{Modifications to Table~\ref{tb:metadata}} & \multirow{2}{*}{Sensitivity} & \multicolumn{2}{c}{Optimal SNR of inspiral} & \multirow{2}{*}{$\Delta\mathcal{A}/\mathcal{A}_{1+2}^{\rm 1g}$} & \multirow{2}{*}{$\Delta\mathcal{A}/\Delta\mathcal{A}_{\rm expected}$$^b$} \\ \cline{4-5}
                                    &                                                            &                              & 1g merger & 2g merger & & \\ \hline
\multirow{2}{*}{Fiducial injection} & \multirow{2}{*}{No modification}                           & O4                           & 20.1 & 28.0 & $0.43_{-0.61}^{+1.38}$ ($87.9\%$)$^c$ & $0.79_{-1.28}^{+2.44}$ \\
                                    &                                                            & O5                           & 48.5 & 67.9 & $0.82_{-0.58}^{+1.81}$ ($98.9\%$) & $1.50_{-1.01}^{+3.29}$ \\ 
\multirow{2}{*}{Asymmetric system}  & $m_1^{\rm 1g}=40M_\odot$, $m_1^{\rm 2g}=48.8M_\odot$,      & O4                           & 15.5 & 18.2 & $-0.26_{-0.45}^{+0.90}$ ($27.3\%$) & $-0.42_{-2.25}^{+1.38}$ \\
                                    & \& $m_2^{\rm 1g}=m_2^{\rm 2g}=10M_\odot$, $a_1^{\rm 2g}=0.52$ & O5                        & 37.4 & 43.4 & $0.06_{-0.47}^{+1.27}$ ($55.4\%$) & $0.14_{-1.52}^{+2.70}$ \\  
\multirow{2}{*}{Precessing system}  & $a_1^{\rm 1g}=a_2^{\rm 1g}=a_2^{\rm 2g}=0.3$, $a_1^{\rm 2g}=0.74$, & O4                   & 20.5 & 20.6 & $0.75_{-0.69}^{+6.99}$ ($96.8\%$) & $1.36_{-1.22}^{+12.78}$ \\
                                    & \& $m_1^{\rm 2g}=47.4M_\odot$, $\theta_1^{\rm 1g}=\theta_1^{\rm 2g}=1.3$ & O5             & 49.3 & 50.7 & $0.40_{-0.44}^{+0.44}$ ($93.7\%$) & $0.73_{-0.83}^{+0.74}$ \\ 
\multirow{2}{*}{Edge-on source}     & \multirow{2}{*}{$\theta_{\rm JN}^{\rm 1g}=\theta_{\rm JN}^{\rm 2g}=1.3$, $d_{\rm L}=500{\rm Mpc}$} & O4    & 22.1 & 19.1 & $0.83_{-0.86}^{+2.24}$ ($94.4\%$) & $1.53_{-1.59}^{+3.97}$ \\
                                    &                                                            & O5                           & 52.7 & 43.5 & $0.68_{-0.28}^{+0.30}$ ($100\%$) & $1.28_{-0.52}^{+0.54}$ \\ \hline
\multicolumn{7}{l}{$^a$ Please note that the fiducial injection also has mild precession and the edge-on source is close to (not exactly) edge-on.} \\
\multicolumn{7}{l}{$^b$ $\Delta\mathcal{A}_{\rm expected}$ means the change of the BH horizon areas calculated with injection values.} \\
\multicolumn{7}{l}{$^c$ The median values and $90\%$ bounds on the fractional change of the BH horizon areas.} \\
\multicolumn{7}{l}{~~~~The percentages of the posteriors that confirm the BH area law are given in parentheses.}                                                  
\end{tabular}
\end{table*}
The intrinsic parameters we have chosen are based on population studies of observed BBH systems \citep{2021arXiv211103634T}. We choose masses in the vicinity of the peaks of the BH mass spectrum and employ nearly aligned and low spins that are consistent with observations. While the selection of other extrinsic parameters is somewhat arbitrary. In addition to the injection configurations in Table~\ref{tb:metadata}, we also carry out more simulations with different injection values of the parameters. We list a few representative tests and their information (e.g., the optimal SNR for the inspiral portion of 1g/2g merger) in Table~\ref{tb:tests}. We perform a series of injections for each test using different noise realizations, i.e., the injected noises are generated with different random seeds. As parameter inference is time consuming, here we only examine five injections (labelled as `Noise 0' -- `Noise 4') into different noise realizations. Unless otherwise noted, all presented results are based on `Noise 0'. The GW data analysis toolkits used in this work include {\sc Bilby} \citep{2019ApJS..241...27A}, {\sc Pycbc} \citep{2019PASP..131b4503B}, and {\sc LalSuite} \citep{lalsuite}. The nest sampling package is {\sc Dynesty} \citep{2020MNRAS.493.3132S}, with the following settings: the stopping criterion (i.e., the evidence tolerance) is `dlogz:0.001', the number of live points is $2000$, and other settings are the same to the default settings adopted by {\sc Bilby}.

\section{Results}
\begin{figure}
\centering
\includegraphics[width=0.48\textwidth]{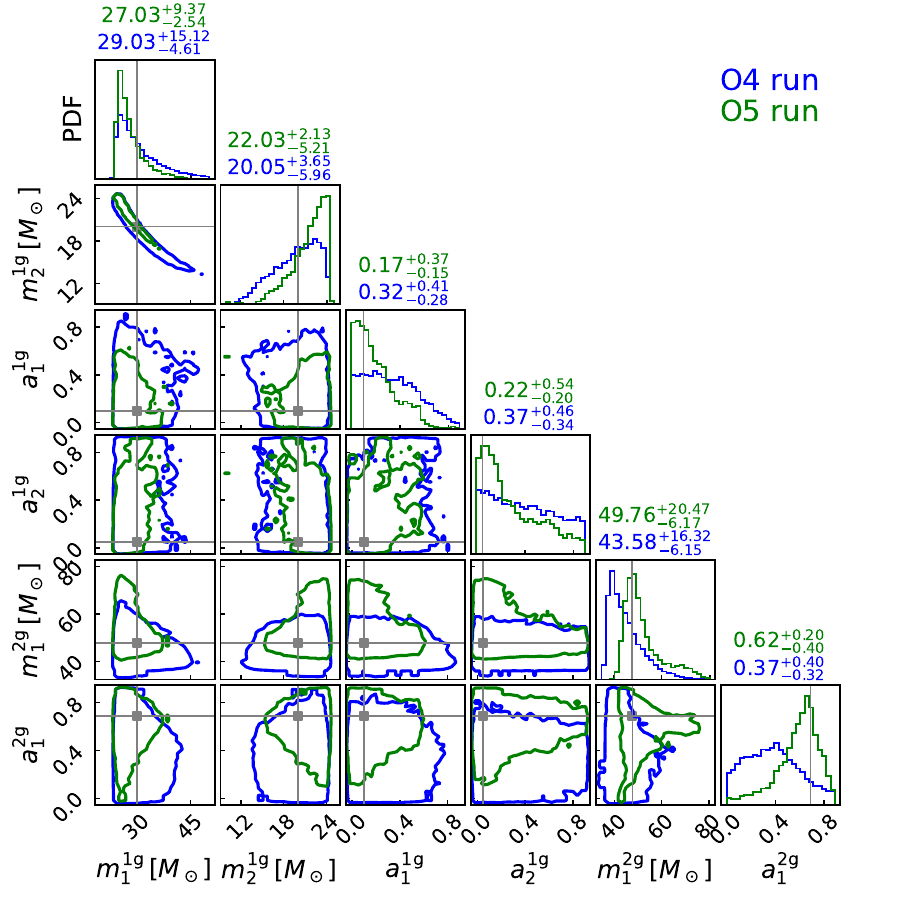}
\caption{Posterior distributions of mass and spin parameters of the 1g and 2g mergers for the fiducial injection. The grey lines represent the injected values. The contours of the joint distributions and the uncertainties reported above the diagonal subfigures are at the $90\%$ credible level. The blue and green colours represent the results obtained with O4 and O5 sensitivities, respectively.}
\label{fig:corner-plots}
\end{figure}
From the posterior distributions (as shown in Fig.~\ref{fig:corner-plots}) of the fiducial injection, we find that the recovered mass properties of BHs in 1g and 2g mergers are more tightly constrained than the spin properties. This is understandable since spin effects appear at higher post-Newtonian orders and the waveform is more sensitive to the effective spin $\chi_{\rm eff}$ rather than the spin magnitude $a_i$. In the case of low SNR (O4), the parameter uncertainties are larger than those obtained with high SNR (O5), as expected. We also notice that some grey points (the injection values) in Fig.~\ref{fig:corner-plots} are on the edge of the $90\%$ contours. This phenomenon was also found in previous studies \citep[e.g.,][]{2015PhRvD..91d2003V, 2023PhRvD.107j3049X}. The inaccurate estimations may be attributed to the following factors: (1) the repercussions of the nearly aligned tilt angle $\theta_{1}^{\rm 2g}$ that has a low prior probability, resulting in the recovered spin magnitudes (partially degenerates with the tilt angle) being shifted from their true values; (2) the contamination from an unexpected noise that makes the signal look like originating from different source parameters. 
\begin{figure*}
\centering
\includegraphics[width=0.48\textwidth]{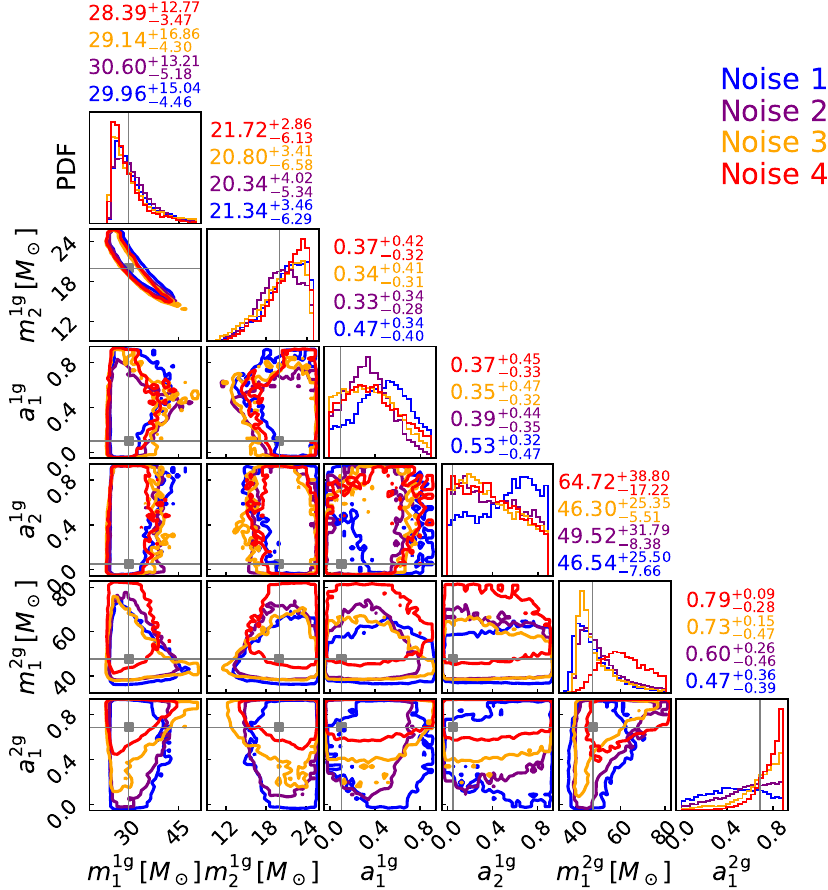}
\includegraphics[width=0.48\textwidth]{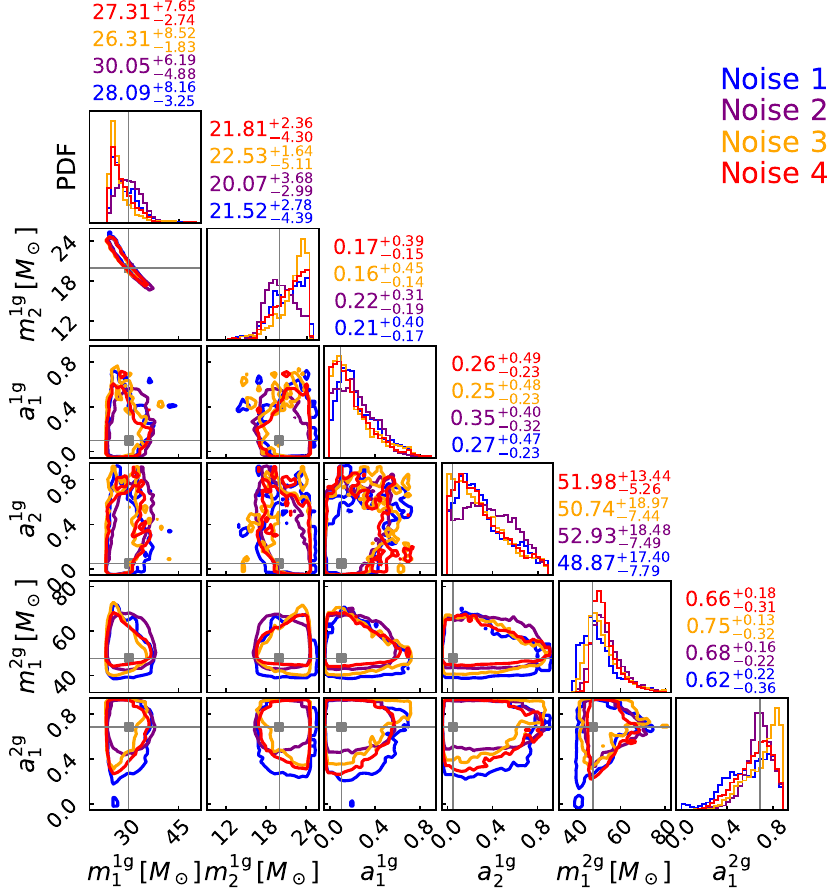}
\caption{Posterior distributions of mass and spin parameters obtained with different noise realizations for the fiducial injection. The panel on the left is for O4 sensitivity, while the one on the right is for O5. The grey lines represent the injected values. The contours of the joint distributions and the uncertainties reported above the diagonal subfigures are at the $90\%$ credible level.}
\label{fig:noise-test}
\end{figure*}
To assess the second factor's impact, we have conducted multiple injections with varying noise realizations. As shown in Fig.~\ref{fig:noise-test}, we find that most parameters are consistently recovered for the majority of injections into different noise realizations. However, the effects of different noise realizations on the parameter estimation results are more pronounced for O4 sensitivity.
\begin{figure}
\centering
\includegraphics[width=0.48\textwidth]{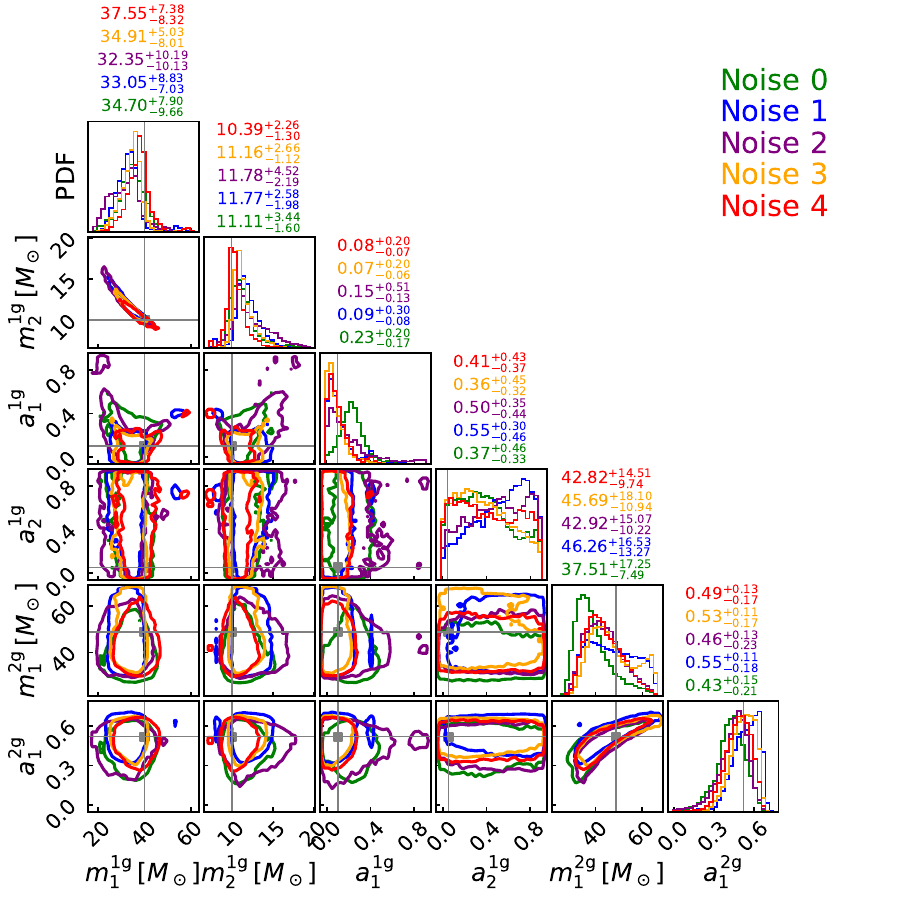}
\caption{Similar to Fig.~\ref{fig:noise-test}, but is the results for the asymmetric system with O5 sensitivity.}
\label{fig:asymmetric-system}
\end{figure}
\begin{figure}
\centering
\includegraphics[width=0.48\textwidth]{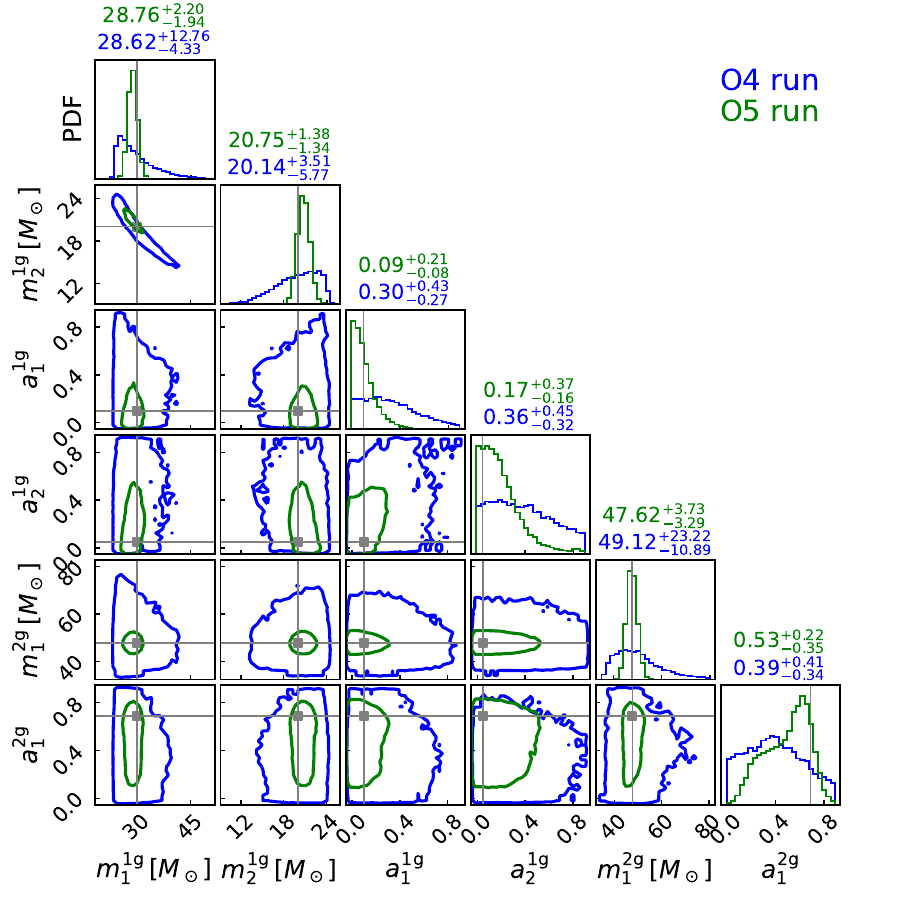}
\caption{Similar to Fig.~\ref{fig:corner-plots}, but is the results for the edge-on source.}
\label{fig:edge-on-source}
\end{figure}
As we know, there are degeneracies present among GW parameters, e.g., the component masses ($m_1$, $m_2$), and the effective spin and mass ratio ($\chi_{\rm eff}$, $q$). These degeneracies may be partly broken for very asymmetric systems (that emit GWs with contributions from higher multipoles) and strong precessing or nearly edge-on systems. We thus have performed a series of additional tests (see Table~\ref{tb:tests}) as we mentioned in the Methods Section. We find that precession effects only have a slight improvement in constraining the intrinsic parameters. In the case of the asymmetric system, the recovered component masses appear to be biased (as shown in Fig.~\ref{fig:asymmetric-system}), and the underlying cause of this bias is not yet known. Additionally, the spin of the secondary component in the first merger remains unconstrained, as the primary component dominates the spin effect in this scenario. While for some edge-on systems with large SNRs, we find that the uncertainties of individual component mass and spin magnitude are significantly reduced (as shown in Fig.~\ref{fig:edge-on-source}).

\begin{figure*}
\centering
\includegraphics[width=0.46\textwidth]{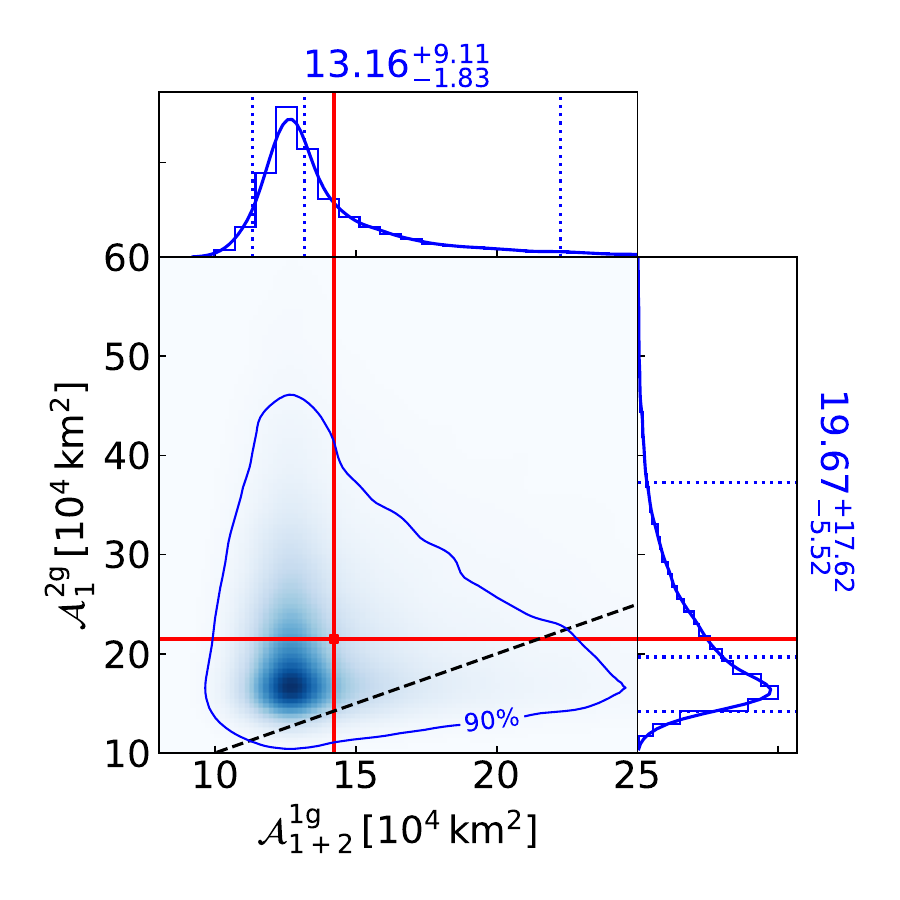}
\includegraphics[width=0.46\textwidth]{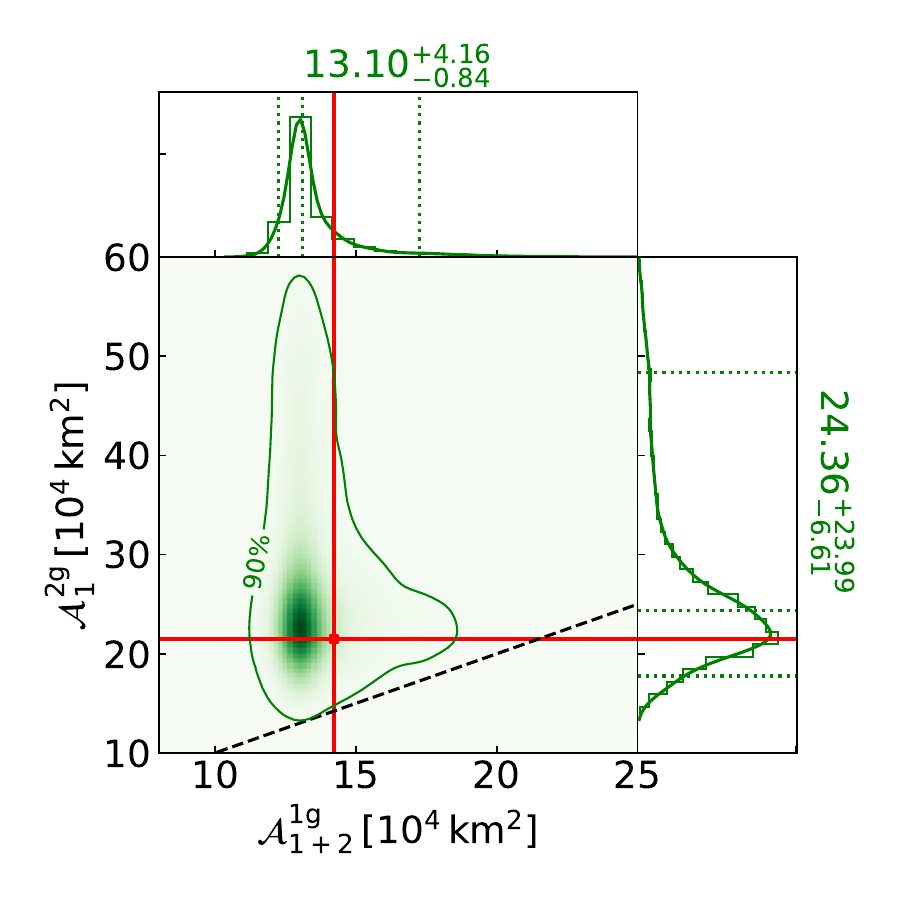}
\includegraphics[width=0.46\textwidth]{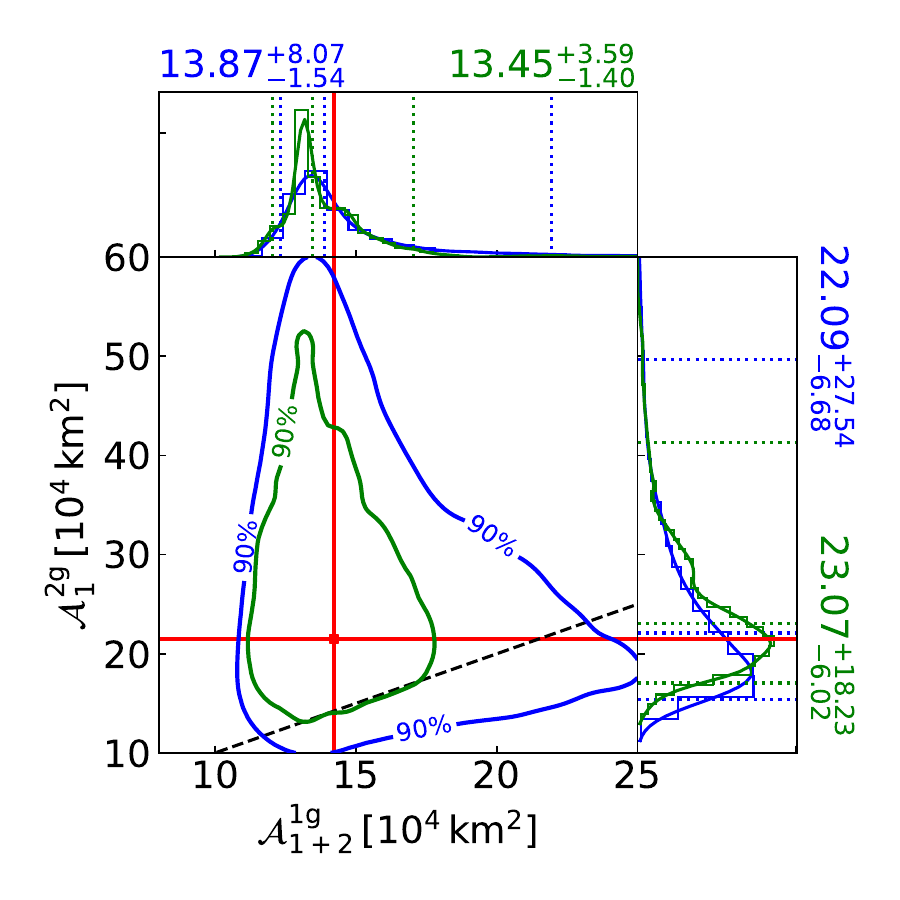}
\includegraphics[width=0.46\textwidth]{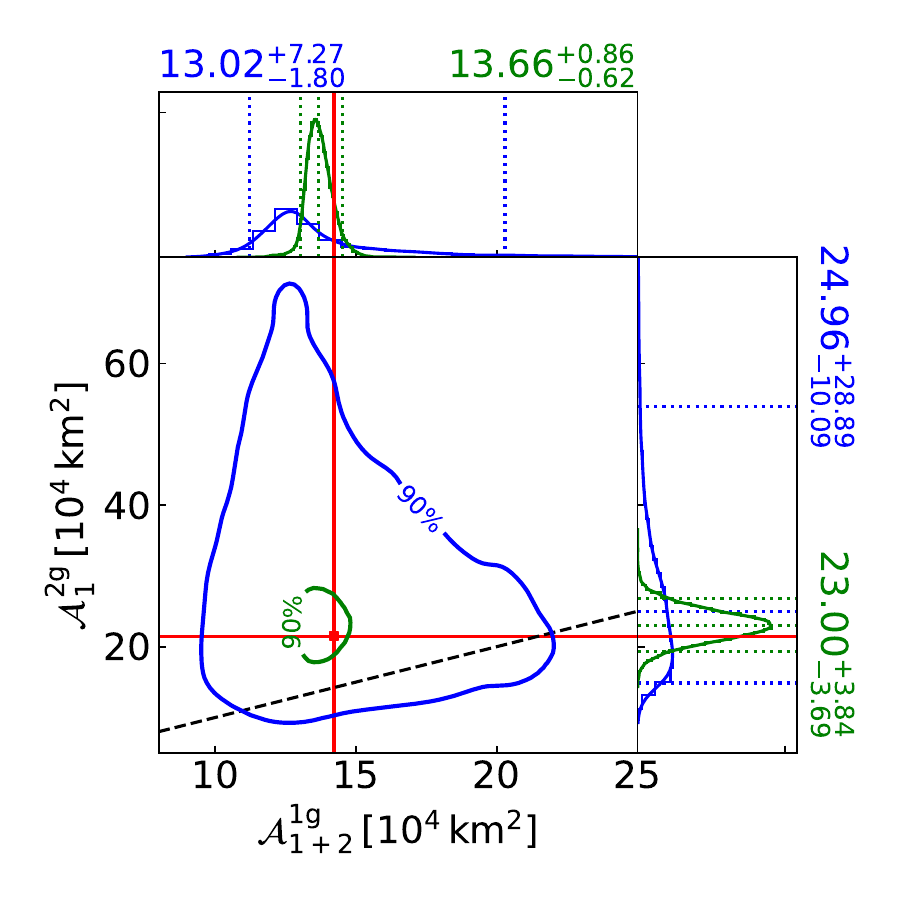}
\caption{Posterior distributions of the total BH area of 1g merger $\mathcal{A}_{1+2}^{\rm 1g}$ and the remnant BH (formed after the 1g merger) area $\mathcal{A}_1^{\rm 2g}$. The red lines in each subfigure represent the injected values. The dotted lines denote the $5\%$, $50\%$, and $95\%$ percentiles. The contours of the joint distributions and the uncertainties of the marginal distributions are at the $90\%$ credible level. The top panels show the results of the fiducial injection for `Noise 0', while the bottom left panel shows the results of the fiducial injection for `Noise 1'. The bottom right panel displays the results of the edge-on source for `Noise 0'. The blue and green colours represent the results obtained with O4 and O5 sensitivities, respectively. The black-dashed line marks $\mathcal{A}_{1+2}^{\rm 1g}=\mathcal{A}_1^{\rm 2g}$.}
\label{fig:area-density-plots}
\end{figure*}
\begin{figure}
\centering
\includegraphics[width=0.48\textwidth]{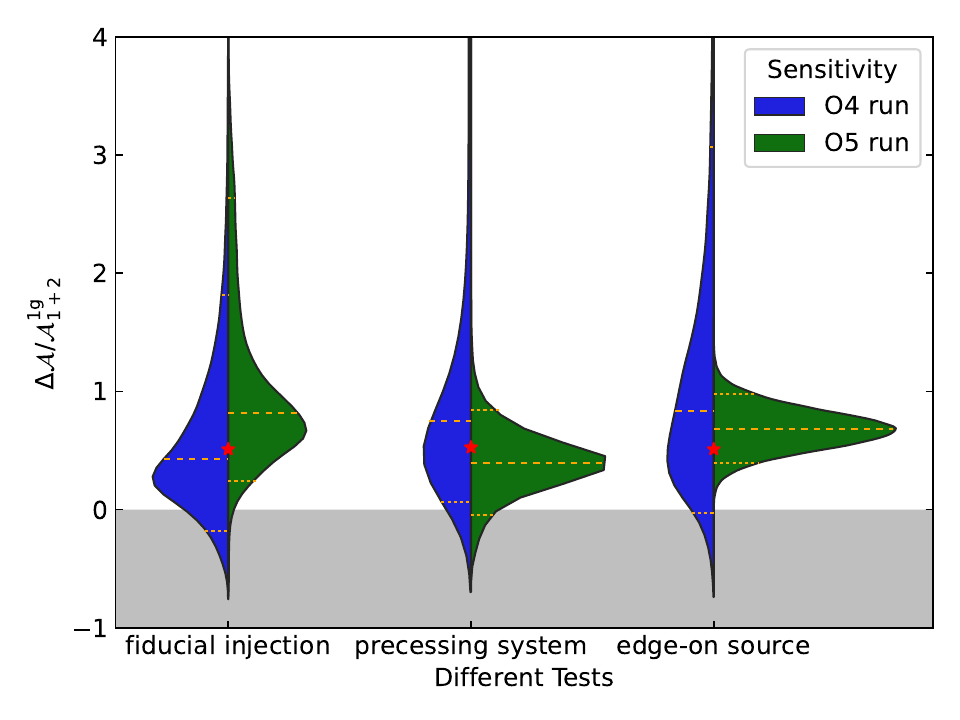}
\caption{Violin plots of fractional change $\Delta\mathcal{A}/\mathcal{A}_{1+2}^{\rm 1g}$ in the horizon area before and after the 1g merger for different tests. The red stars represent the fractional changes calculated with injection values. The dashed lines denote the $5\%$, $50\%$, and $95\%$ percentiles.}
\label{fig:area-violin-plots}
\end{figure}

The Kerr BH horizon areas can be calculated via 
\begin{equation}
\mathcal{A}(m,a)=8\pi (Gm/c^2)^2 (1+\sqrt{1-a^2}),
\label{eq:farea-m-a}
\end{equation}
where $m$ is the gravitational mass and $a$ is the dimensionless spin magnitude. After transforming the posterior samples of mass and spin to the BH areas, we find that the total BH area of 1g merger $\mathcal{A}_{1+2}^{\rm 1g}=\mathcal{A}(m_1^{\rm 1g},a_1^{\rm 1g})+\mathcal{A}(m_2^{\rm 1g},a_2^{\rm 1g})$ and the remnant BH (formed after the 1g merger) area $\mathcal{A}_1^{\rm 2g}=\mathcal{A}(m_1^{\rm 2g},a_1^{\rm 2g})$ are well recovered (as shown in Fig.~\ref{fig:area-density-plots}; though the posterior distribution region with maximum probability does not perfectly cover the true values) compared to the mass and spin properties. The reason why the recovery of BH area is better than mass and spin properties is that the total BH area of 1g merger $\mathcal{A}_{1+2}^{\rm 1g} \propto m_{\rm T}^2-2(\mathcal{M}^5m_{\rm T})^{1/3}$ is mainly dependent on the chirp mass $\mathcal{M}$ or total mass $m_{\rm T}$ and the deviation of BH area caused by spin magnitude is relatively small for low spin injection since $\partial \ln \mathcal{A}/\partial a = -a/(1-a^2+\sqrt{1-a^2})$. We can notice that $\mathcal{A}_{1+2}^{\rm 1g}$ is tightly constrained for both O4 and O5 sensitivities, while $\mathcal{A}_1^{\rm 2g}$ exhibits a broader distribution since it is more sensitive to the recovered mass ratio distribution compared to $\mathcal{A}_{1+2}^{\rm 1g}$. For our fiducial injection with `Noise 0', the uncertainty of $\mathcal{A}_1^{\rm 2g}$ obtained with O5 sensitivity appears to be larger than that obtained with O4, which is not expected. However, after examining the results obtained with other noise realizations (for example, in the bottom left panel of Fig.~\ref{fig:area-density-plots}), we conclude that this is only an exceptional occurrence. The $90\%$ contour region of BH areas obtained with a high SNR (O5) is shrunk, and a majority of posteriors are above the black-dashed line $\mathcal{A}_{1+2}^{\rm 1g}=\mathcal{A}_1^{\rm 2g}$, which shows good consistency with the prediction of BH area law. To evaluate what credible level one can achieve in the hierarchical triple merger scenario, we calculate the fractional change in the horizon area before and after the 1g merger. The results of $\Delta\mathcal{A}/\mathcal{A}_{1+2}^{\rm 1g}$ ($\Delta\mathcal{A} \equiv \mathcal{A}_1^{\rm 2g}-\mathcal{A}_{1+2}^{\rm 1g}$) are shown in Fig.~\ref{fig:area-violin-plots}. Apart from the asymmetric system, which yields poor results (as shown in Table~\ref{tb:tests}) due to the biased recovery of intrinsic parameters, the other tests exhibit favourable outcomes in testing the area law. The probabilities of fractional changes that self-consistently confirm the BH area law are $87.9\%$ (O4) and $98.9\%$ (O5) for the fiducial injection. Moreover, for edge-on sources detected in the O5 run, this probability even reaches $\sim100\%$. Therefore, using hierarchical triple mergers to test the BH area law is promising, which is complementary to previous methods \citep{2018PhRvD..97l4069C, 2021PhRvL.127a1103I, 2022PhRvD.105f4042K}.

\section{Discussion}
\begin{figure}
\centering
\includegraphics[width=0.48\textwidth]{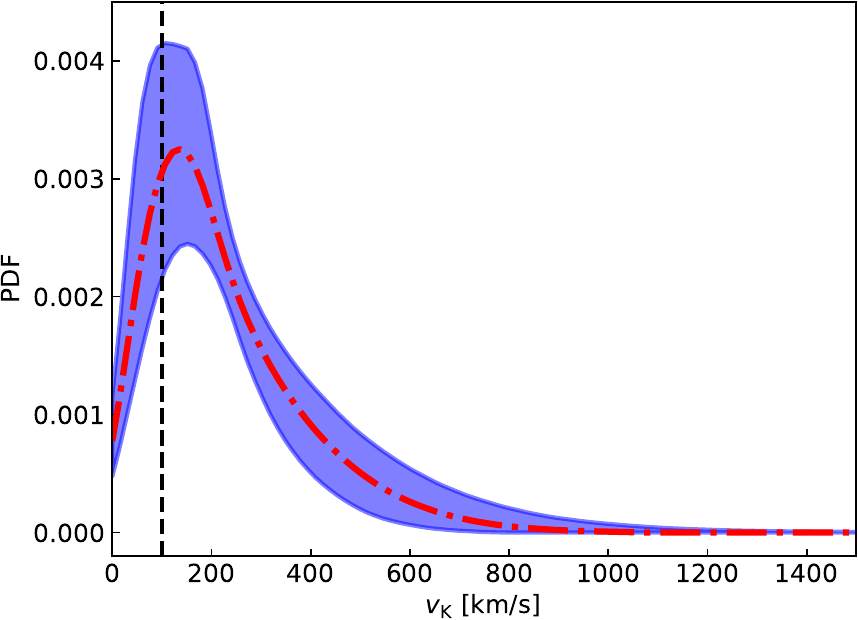}
\caption{Distribution of the recoil velocities of 1g mergers inferred with the sub-population (i.e., `dynamical 1G+1G') obtained by \citet{2022ApJ...941L..39W}. The blue area represents the $90\%$ credible region.}
\label{fig:kick-velocity}
\end{figure}

One of the most uncertain factors for the hierarchical triple mergers is their merger rate, which is influenced by their formation rate and birth environments, the recoil kicks of 1g mergers, and the initial configuration of the system at its formation. The data analyses and population studies on the GW events detected by aLIGO/AdV indicate that hierarchical (triple) mergers are possible \citep{2021ApJ...907L..48V, 2021ApJ...907L..19V, 2021NatAs...5..749G, 2022PhR...955....1M}. \citet{2019MNRAS.482...30S} found that the possibility for observing both 1g/2g mergers within a time-scale of $\mathcal{O}(\rm yr)$ is notable (here we interpret `notable' as $10\%$) only for co-planar interactions with low GW kicks ($v_{\rm K} \lesssim\,10-100\,\rm km\,s^{-1}$). Such co-planar interactions may occur in the environments of active galactic nuclei (AGNs) discs, which may contribute a fraction to the observed BBHs \citep{2019PhRvL.123r1101Y, 2022Natur.603..237S, 2021ApJ...920L..42G, 2023arXiv230302973L}. Recently, \citet{2022ApJ...941L..39W} found that $\sim 15\%$ of the underlying astrophysical BBH mergers could form through dynamical processes, in which a fraction of $\sim 10\%$ mergers contain 2g BHs. After considering the selection effect, the fraction containing 2g BHs in the {\it detected} BBHs is about $16\%$. We use a sub-population of BBHs (i.e., the `dynamical 1G+1G' population that has small spin magnitude) obtained by \citet{2022ApJ...941L..39W} to infer the distribution of recoil velocities (as shown in Fig.~\ref{fig:kick-velocity}), and we find that the fraction of 1g merger remnants with low recoil velocities ($v_{\rm K}\lesssim 100\,\rm km\,s^{-1}$) is ${20}_{-6}^{+10}\%$ at the $90\%$ credible level. This value is larger than the results of previous studies \citep{2021ApJ...914L..18D, 2022PhRvL.128c1101V} because the sub-population we used only includes 1g BHs.

In the optimistic case (i.e., these 2g BHs mainly originate from AGNs discs), the hierarchical triple mergers with a 2g merger time-scale $\tau_{\rm 2g}<\mathcal{O}(\rm yr)$ may occupy 
\begin{equation}
\begin{aligned}
&~p({\rm both~mergers~are~found~in~the~observed~BBHs})\\
&\approx p({\rm 2g~detected})p({\rm 1g~detected}|{\rm 2g~detected}, \tau_{\rm 2g}<\mathcal{O}({\rm yr})) \\
&~~~~~p(\tau_{\rm 2g}<\mathcal{O}({\rm yr}) | {\rm AGN, low~kick})p({\rm AGN})p({\rm low~kick})\\
&=0.16\times0.46\times0.1\times1.0\times0.2 \sim 0.1\%
\end{aligned}
\label{eq:p1g2g}
\end{equation}
of the detected BBHs (please note that this is only a rough estimate, which should be calibrated with future observations). In the above equation, we have assumed independence of the probabilities in the second and third lines. Once we observe a 2g BBH merger, there must have a progenitor 1g merger which may have happened before the 2g merger on a time-scale ranging from $<\mathcal{O}({\rm yr})$ to $\sim{\rm Gyr}$. Given the condition that the 2g merger is detected, the probability of finding the progenitor 1g merger, i.e., $p({\rm 1g~detected}|{\rm 2g~detected})$ is mainly determined by two factors; one is the portion of hierarchical triple systems with a 2g merger time-scale $<\mathcal{O}(\rm yr)$ (described by the third line of Equation \ref{eq:p1g2g}), and the other is the relative detectable volumes of 1g and 2g mergers, i.e., $p({\rm 1g~detected}|{\rm 2g~detected}, \tau_{\rm 2g}<\mathcal{O}({\rm yr})) \approx \mathcal{V}_{\rm 1g}/\mathcal{V}_{\rm 2g}$, which approximates to $(\mathcal{M}^{\rm 1g}/\mathcal{M}^{\rm 2g})^{5/2}$ (please note that this is also a rough estimate and can be derived using the formula from, e.g., \citealt{2021CQGra..38e5010C}). Since the population properties of hierarchical triple systems are unclear currently, we set this value to $0.46$ for the equal mass case, i.e., $m_1^{\rm 1g}=m_2^{\rm 1g}=m_2^{\rm 2g}$.

Assuming that the O4 (O5) run will last for $1\,\rm yr$ ($2\,\rm yr$) and the reachable range for O4 (O5) is $\sim1.5$ ($\sim2.5$) times that of O3 \citep{2018LRR....21....3A}, there will be $\mathcal{O}(3000)$ BBH mergers being detected. Therefore, catching a pair of such events at the end of the O5 run is possible. Besides, it is feasible to identify these kinds of merger events by using their characteristics, e.g., (1) the total mass of the 1g merger is close to one of the BHs in the 2g merger, (2) the dimensionless spin magnitude for one of the 2g merger components is close to $0.7$, and (3) the 1g/2g mergers share the same position \citep{2021ApJ...907L..48V}. Though lensing GW signals may also have the characteristic of sharing the same sky location, their other characteristics are different from the hierarchical triple merger scenario. Finally, please note that the orbit separation between ${\rm BH}_1$\textendash{}${\rm BH}_2$ and ${\rm BH}_3$ is $>\mathcal{O}(10^4\,{\rm km})$ when the time-scale between the first and second mergers is $>\mathcal{O}(\rm hr)$, and ${\rm BH}_3$ (see Fig.~\ref{fig:hierarchical-triple-merger}) will not influence the inspiral signal of the first merger. This is because (1) the inspiral signal of the first merger entering the LIGO band is short (less than a few seconds), and the third distant BH (i.e., ${ \rm BH}_3$, whose gravitational field felt by ${ \rm BH}_1$\textendash{}${ \rm BH}_2$ is weak and stationary) will not significantly impact the dynamics of the binary system within a short period (although secular gravitational interaction may have a negligible effect); (2) the GWs emitted by the inspiral of ${ \rm BH}_3$ with the binary system are not detectable in the LIGO band and will not contaminate the inspiral signal of the first merger.

We would like to note some caveats on our results due to the assumptions and uncertainties in the investigation. It may be challenging for the application of testing the BH area law with the hierarchical triple mergers if the mass ratio of the 2g merger is close to one and both of the components of the 2g merger have $a_i \sim 0.7$ (the remnant of the 1g merger is indistinguishable in this situation). We have ignored some measurement errors that would appear in the real data analysis. First, we did not consider the detector calibration errors that introduce uncertainties regarding the strain's scale and phase. However, \citet{2016PhRvL.116x1102A} reported that the calibration errors would influence the sky localization but has little effect on the estimates of intrinsic parameters. Secondly, we fixed the PSDs as those used in injections when inferring the GW parameters. A more robust but complicated procedure is parameterizing the PSD estimation uncertainties in the likelihood function \citep{2015PhRvD..91d2003V, 2019PhRvD.100j4004C, 2020PhRvD.102b3008B}. Thirdly, we did not consider the systematic bias caused by the waveform templates. The short-time hierarchical triple mergers were expected to possess non-negligible eccentricity \citep{2019MNRAS.482...30S}. The {\sc IMRPhenomXPHM} used in this work does not consider the orbital eccentricity, which may produce bias when applied to real data analysis \citep{2023PhRvD.107b4009B}. Though comprehensive waveform templates including the effects of precession, higher-order multipole moments, and orbital eccentricity, are not yet available, work to improve the waveform model is well underway \citep{2019PhRvD.100j4036A}. And a thorough assessment of the waveform systematics is beyond the scope of this work. Fourthly, only a small number of injections into different noise realizations are examined in this work. More simulations are needed to investigate the statistical properties of our method.

\section{Summary}
In this work, we propose that the hierarchical triple merger scenario can be used to test Hawking's area theorem. Such a test only needs the inspiral portions of the 1g and 2g mergers which can independently measure the properties of the progenitor BHs and the remnant BH. By recovering the parameters of simulated events with inspiral-only data analyses, we find that the BH horizon areas before/after the 1g merger can be well constrained. The change in the total horizon area then permits a test of the BH area law. Under some plausible configurations of injections, we find that the law can be confirmed with $\sim 88\%-99\%$ credibility for the majority of injections. With the upgrade of aLIGO/AdV and the participation of KAGRA \citep{2013PhRvD..88d3007A} and LIGO-India \citep{2022CQGra..39b5004S}, thousands of BBH events will be detected collaboratively by the detector network in the future. Therefore, if the hierarchical triple mergers contribute a non-negligible fraction ($\gtrsim 0.1\%$) to the detected BBHs, our method can facilitate the test of BH area law and is complementary to previous studies \citep{2021PhRvL.127a1103I, 2022PhRvD.105f4042K}. In addition, by combining the analysis of the two inspiral signals along with the ringdown of the 1g merger and/or by incorporating the data obtained from future spaceborne GW detectors into a multiband analysis, we expect to further enhance the effectiveness of our approach in testing the area law.

\section*{Data Availability}
The derived data generated in this research will be shared upon reasonable request to the corresponding author. The noise curves used in this work are taken from \href{https://dcc.ligo.org/LIGO-T2000012/public}{https://dcc.ligo.org/LIGO-T2000012/public}. \\
{\it Software}: {\sc Bilby} \citep[version 1.4.1;][]{2019ApJS..241...27A}, {\sc Dynesty} \citep[version 2.1.0;][]{2020MNRAS.493.3132S}, and {\sc PyCBC} \citep[version 2.0.6;][]{2019PASP..131b4503B}.

\section*{Acknowledgements}
We thank the anonymous referee for the helpful suggestions. We also thank YF Wang and HT Wang for their valuable discussions. SP Tang acknowledges funding from the Special Research Assistant Project of CAS. This work was supported in part by NSFC under grants no. 12233011, no. 11921003, no. 11933010, and no. 12073080.

\bibliographystyle{mnras}
\bibliography{bibtex}

\begin{thebibliography}{}
\makeatletter
\relax
\def\mn@urlcharsother{\let\do\@makeother \do\$\do\&\do\#\do\^\do\_\do\%\do\~}
\def\mn@doi{\begingroup\mn@urlcharsother \@ifnextchar [ {\mn@doi@}
  {\mn@doi@[]}}
\def\mn@doi@[#1]#2{\def\@tempa{#1}\ifx\@tempa\@empty \href
  {http://dx.doi.org/#2} {doi:#2}\else \href {http://dx.doi.org/#2} {#1}\fi
  \endgroup}
\def\mn@eprint#1#2{\mn@eprint@#1:#2::\@nil}
\def\mn@eprint@arXiv#1{\href {http://arxiv.org/abs/#1} {{\tt arXiv:#1}}}
\def\mn@eprint@dblp#1{\href {http://dblp.uni-trier.de/rec/bibtex/#1.xml}
  {dblp:#1}}
\def\mn@eprint@#1:#2:#3:#4\@nil{\def\@tempa {#1}\def\@tempb {#2}\def\@tempc
  {#3}\ifx \@tempc \@empty \let \@tempc \@tempb \let \@tempb \@tempa \fi \ifx
  \@tempb \@empty \def\@tempb {arXiv}\fi \@ifundefined
  {mn@eprint@\@tempb}{\@tempb:\@tempc}{\expandafter \expandafter \csname
  mn@eprint@\@tempb\endcsname \expandafter{\@tempc}}}

\bibitem[\protect\citeauthoryear{{Abbott}, {Abbott}, {Abbott}  et~al.}{{Abbott}
  et~al.}{2016a}]{2016PhRvL.116f1102A}
{Abbott} B.~P.,  {Abbott} R.,  {Abbott} T.~D.,   et~al., 2016a, \mn@doi [\prl]
  {10.1103/PhysRevLett.116.061102}, \href
  {https://ui.adsabs.harvard.edu/abs/2016PhRvL.116f1102A} {116, 061102}

\bibitem[\protect\citeauthoryear{{Abbott}, {Abbott}, {Abbott}  et~al.}{{Abbott}
  et~al.}{2016b}]{2016PhRvL.116x1102A}
{Abbott} B.~P.,  {Abbott} R.,  {Abbott} T.~D.,   et~al., 2016b, \mn@doi [\prl]
  {10.1103/PhysRevLett.116.241102}, \href
  {https://ui.adsabs.harvard.edu/abs/2016PhRvL.116x1102A} {116, 241102}

\bibitem[\protect\citeauthoryear{{Abbott}, {Abbott}, {Abbott}  et~al.}{{Abbott}
  et~al.}{2018}]{2018LRR....21....3A}
{Abbott} B.~P.,  {Abbott} R.,  {Abbott} T.~D.,   et~al., 2018, \mn@doi [Living
  Reviews in Relativity] {10.1007/s41114-018-0012-9}, \href
  {https://ui.adsabs.harvard.edu/abs/2018LRR....21....3A} {21, 3}

\bibitem[\protect\citeauthoryear{{Abbott}, {Abbott}, {Abbott}  et~al.}{{Abbott}
  et~al.}{2019}]{2019PhRvD.100j4036A}
{Abbott} B.~P.,  {Abbott} R.,  {Abbott} T.~D.,   et~al., 2019, \mn@doi [\prd]
  {10.1103/PhysRevD.100.104036}, \href
  {https://ui.adsabs.harvard.edu/abs/2019PhRvD.100j4036A} {100, 104036}

\bibitem[\protect\citeauthoryear{{Abbott}, {Abbott}, {Abbott}  et~al.}{{Abbott}
  et~al.}{2020}]{2020CQGra..37e5002A}
{Abbott} B.~P.,  {Abbott} R.,  {Abbott} T.~D.,   et~al., 2020, \mn@doi
  [Classical and Quantum Gravity] {10.1088/1361-6382/ab685e}, \href
  {https://ui.adsabs.harvard.edu/abs/2020CQGra..37e5002A} {37, 055002}

\bibitem[\protect\citeauthoryear{{Abbott}, {Abbott}, {Acernese}
  et~al.}{{Abbott} et~al.}{2021}]{2021arXiv211103634T}
{Abbott} R.,  {Abbott} T.~D.,  {Acernese} F.,   et~al., 2021, \mn@doi [arXiv
  e-prints] {10.48550/arXiv.2111.03634}, \href
  {https://ui.adsabs.harvard.edu/abs/2021arXiv211103634T} {p. arXiv:2111.03634}

\bibitem[\protect\citeauthoryear{{Acernese}, {Agathos}, {Agatsuma}
  et~al.}{{Acernese} et~al.}{2015}]{2015CQGra..32b4001A}
{Acernese} F.,  {Agathos} M.,  {Agatsuma} K.,   et~al., 2015, \mn@doi
  [Classical and Quantum Gravity] {10.1088/0264-9381/32/2/024001}, \href
  {https://ui.adsabs.harvard.edu/abs/2015CQGra..32b4001A} {32, 024001}

\bibitem[\protect\citeauthoryear{{Ashton}, {H{\"u}bner}, {Lasky}
  et~al.}{{Ashton} et~al.}{2019}]{2019ApJS..241...27A}
{Ashton} G.,  {H{\"u}bner} M.,  {Lasky} P.~D.,   et~al., 2019, \mn@doi [\apjs]
  {10.3847/1538-4365/ab06fc}, \href
  {https://ui.adsabs.harvard.edu/abs/2019ApJS..241...27A} {241, 27}

\bibitem[\protect\citeauthoryear{{Aso}, {Michimura}, {Somiya}  et~al.}{{Aso}
  et~al.}{2013}]{2013PhRvD..88d3007A}
{Aso} Y.,  {Michimura} Y.,  {Somiya} K.,   et~al., 2013, \mn@doi [\prd]
  {10.1103/PhysRevD.88.043007}, \href
  {https://ui.adsabs.harvard.edu/abs/2013PhRvD..88d3007A} {88, 043007}

\bibitem[\protect\citeauthoryear{{Bhat}, {Saini}, {Favata}  \& {Arun}}{{Bhat}
  et~al.}{2023}]{2023PhRvD.107b4009B}
{Bhat} S.~A.,  {Saini} P.,  {Favata} M.,   {Arun} K.~G.,  2023, \mn@doi [\prd]
  {10.1103/PhysRevD.107.024009}, \href
  {https://ui.adsabs.harvard.edu/abs/2023PhRvD.107b4009B} {107, 024009}

\bibitem[\protect\citeauthoryear{{Biscoveanu}, {Haster}, {Vitale}
  et~al.}{{Biscoveanu} et~al.}{2020}]{2020PhRvD.102b3008B}
{Biscoveanu} S.,  {Haster} C.-J.,  {Vitale} S.,   et~al., 2020, \mn@doi [\prd]
  {10.1103/PhysRevD.102.023008}, \href
  {https://ui.adsabs.harvard.edu/abs/2020PhRvD.102b3008B} {102, 023008}

\bibitem[\protect\citeauthoryear{{Biwer}, {Capano}  et~al.}{{Biwer}
  et~al.}{2019}]{2019PASP..131b4503B}
{Biwer} C.~M.,  {Capano} C.~D.,   et~al., 2019, \mn@doi [\pasp]
  {10.1088/1538-3873/aaef0b}, \href
  {https://ui.adsabs.harvard.edu/abs/2019PASP..131b4503B} {131, 024503}

\bibitem[\protect\citeauthoryear{{Cabero}, {Nielsen}, {Lundgren}  \&
  {Capano}}{{Cabero} et~al.}{2017}]{2017PhRvD..95f4016C}
{Cabero} M.,  {Nielsen} A.~B.,  {Lundgren} A.~P.,   {Capano} C.~D.,  2017,
  \mn@doi [\prd] {10.1103/PhysRevD.95.064016}, \href
  {https://ui.adsabs.harvard.edu/abs/2017PhRvD..95f4016C} {95, 064016}

\bibitem[\protect\citeauthoryear{{Cabero}, {Capano}, {Fischer-Birnholtz}
  et~al.}{{Cabero} et~al.}{2018}]{2018PhRvD..97l4069C}
{Cabero} M.,  {Capano} C.~D.,  {Fischer-Birnholtz} O.,   et~al., 2018, \mn@doi
  [\prd] {10.1103/PhysRevD.97.124069}, \href
  {https://ui.adsabs.harvard.edu/abs/2018PhRvD..97l4069C} {97, 124069}

\bibitem[\protect\citeauthoryear{{Capano} et~al.,}{{Capano}
  et~al.}{2021}]{2021arXiv210505238C}
{Capano} C.~D.,  et~al., 2021, \mn@doi [arXiv e-prints]
  {10.48550/arXiv.2105.05238}, \href
  {https://ui.adsabs.harvard.edu/abs/2021arXiv210505238C} {p. arXiv:2105.05238}

\bibitem[\protect\citeauthoryear{{Chatziioannou}, {Haster}, {Littenberg}
  et~al.}{{Chatziioannou} et~al.}{2019}]{2019PhRvD.100j4004C}
{Chatziioannou} K.,  {Haster} C.-J.,  {Littenberg} T.~B.,   et~al., 2019,
  \mn@doi [\prd] {10.1103/PhysRevD.100.104004}, \href
  {https://ui.adsabs.harvard.edu/abs/2019PhRvD.100j4004C} {100, 104004}

\bibitem[\protect\citeauthoryear{{Chen}, {Holz}, {Miller}, {Evans}, {Vitale}
  \& {Creighton}}{{Chen} et~al.}{2021}]{2021CQGra..38e5010C}
{Chen} H.-Y.,  {Holz} D.~E.,  {Miller} J.,  {Evans} M.,  {Vitale} S.,
  {Creighton} J.,  2021, \mn@doi [Classical and Quantum Gravity]
  {10.1088/1361-6382/abd594}, \href
  {https://ui.adsabs.harvard.edu/abs/2021CQGra..38e5010C} {38, 055010}

\bibitem[\protect\citeauthoryear{{Cornish} \& {Littenberg}}{{Cornish} \&
  {Littenberg}}{2015}]{2015CQGra..32m5012C}
{Cornish} N.~J.,  {Littenberg} T.~B.,  2015, \mn@doi [Classical and Quantum
  Gravity] {10.1088/0264-9381/32/13/135012}, \href
  {https://ui.adsabs.harvard.edu/abs/2015CQGra..32m5012C} {32, 135012}

\bibitem[\protect\citeauthoryear{{Doctor}, {Farr}  \& {Holz}}{{Doctor}
  et~al.}{2021}]{2021ApJ...914L..18D}
{Doctor} Z.,  {Farr} B.,   {Holz} D.~E.,  2021, \mn@doi [\apjl]
  {10.3847/2041-8213/ac0334}, \href
  {https://ui.adsabs.harvard.edu/abs/2021ApJ...914L..18D} {914, L18}

\bibitem[\protect\citeauthoryear{{Gayathri}, {Yang}, {Tagawa}, {Haiman}  \&
  {Bartos}}{{Gayathri} et~al.}{2021}]{2021ApJ...920L..42G}
{Gayathri} V.,  {Yang} Y.,  {Tagawa} H.,  {Haiman} Z.,   {Bartos} I.,  2021,
  \mn@doi [\apjl] {10.3847/2041-8213/ac2cc1}, \href
  {https://ui.adsabs.harvard.edu/abs/2021ApJ...920L..42G} {920, L42}

\bibitem[\protect\citeauthoryear{{Gerosa} \& {Fishbach}}{{Gerosa} \&
  {Fishbach}}{2021}]{2021NatAs...5..749G}
{Gerosa} D.,  {Fishbach} M.,  2021, \mn@doi [Nature Astronomy]
  {10.1038/s41550-021-01398-w}, \href
  {https://ui.adsabs.harvard.edu/abs/2021NatAs...5..749G} {5, 749}

\bibitem[\protect\citeauthoryear{{Ghosh}, {Johnson-McDaniel}, {Ghosh}
  et~al.}{{Ghosh} et~al.}{2018}]{2018CQGra..35a4002G}
{Ghosh} A.,  {Johnson-McDaniel} N.~K.,  {Ghosh} A.,   et~al., 2018, \mn@doi
  [Classical and Quantum Gravity] {10.1088/1361-6382/aa972e}, \href
  {https://ui.adsabs.harvard.edu/abs/2018CQGra..35a4002G} {35, 014002}

\bibitem[\protect\citeauthoryear{{Giudice}, {McCullough}  \&
  {Urbano}}{{Giudice} et~al.}{2016}]{2016JCAP...10..001G}
{Giudice} G.~F.,  {McCullough} M.,   {Urbano} A.,  2016, \mn@doi [\jcap]
  {10.1088/1475-7516/2016/10/001}, \href
  {https://ui.adsabs.harvard.edu/abs/2016JCAP...10..001G} {2016, 001}

\bibitem[\protect\citeauthoryear{{Hawking}}{{Hawking}}{1971}]{1971PhRvL..26.1344H}
{Hawking} S.~W.,  1971, \mn@doi [\prl] {10.1103/PhysRevLett.26.1344}, \href
  {https://ui.adsabs.harvard.edu/abs/1971PhRvL..26.1344H} {26, 1344}

\bibitem[\protect\citeauthoryear{{Hughes} \& {Menou}}{{Hughes} \&
  {Menou}}{2005}]{2005ApJ...623..689H}
{Hughes} S.~A.,  {Menou} K.,  2005, \mn@doi [\apj] {10.1086/428826}, \href
  {https://ui.adsabs.harvard.edu/abs/2005ApJ...623..689H} {623, 689}

\bibitem[\protect\citeauthoryear{{Isi}, {Farr}, {Giesler}  et~al.}{{Isi}
  et~al.}{2021}]{2021PhRvL.127a1103I}
{Isi} M.,  {Farr} W.~M.,  {Giesler} M.,   et~al., 2021, \mn@doi [\prl]
  {10.1103/PhysRevLett.127.011103}, \href
  {https://ui.adsabs.harvard.edu/abs/2021PhRvL.127a1103I} {127, 011103}

\bibitem[\protect\citeauthoryear{{Kastha}, {Capano}, {Westerweck}
  et~al.}{{Kastha} et~al.}{2022}]{2022PhRvD.105f4042K}
{Kastha} S.,  {Capano} C.~D.,  {Westerweck} J.,   et~al., 2022, \mn@doi [\prd]
  {10.1103/PhysRevD.105.064042}, \href
  {https://ui.adsabs.harvard.edu/abs/2022PhRvD.105f4042K} {105, 064042}

\bibitem[\protect\citeauthoryear{{Knee}, {McIver}  \& {Cabero}}{{Knee}
  et~al.}{2022}]{2022ApJ...928...21K}
{Knee} A.~M.,  {McIver} J.,   {Cabero} M.,  2022, \mn@doi [\apj]
  {10.3847/1538-4357/ac48f5}, \href
  {https://ui.adsabs.harvard.edu/abs/2022ApJ...928...21K} {928, 21}

\bibitem[\protect\citeauthoryear{{LIGO Scientific Collaboration}}{{LIGO
  Scientific Collaboration}}{2015}]{2015CQGra..32g4001L}
{LIGO Scientific Collaboration} 2015, \mn@doi [Classical and Quantum Gravity]
  {10.1088/0264-9381/32/7/074001}, \href
  {https://ui.adsabs.harvard.edu/abs/2015CQGra..32g4001L} {32, 074001}

\bibitem[\protect\citeauthoryear{{LIGO Scientific Collaboration}}{{LIGO
  Scientific Collaboration}}{2018}]{lalsuite}
{LIGO Scientific Collaboration} 2018, {LIGO} {A}lgorithm {L}ibrary -
  {LALS}uite, {free software (GPL)}, \mn@doi{10.7935/GT1W-FZ16}

\bibitem[\protect\citeauthoryear{{Li}, {Wang}, {Tang}  \& {Fan}}{{Li}
  et~al.}{2023}]{2023arXiv230302973L}
{Li} Y.-J.,  {Wang} Y.-Z.,  {Tang} S.-P.,   {Fan} Y.-Z.,  2023, \mn@doi [arXiv
  e-prints] {10.48550/arXiv.2303.02973}, \href
  {https://ui.adsabs.harvard.edu/abs/2023arXiv230302973L} {p. arXiv:2303.02973}

\bibitem[\protect\citeauthoryear{{Littenberg} \& {Cornish}}{{Littenberg} \&
  {Cornish}}{2015}]{2015PhRvD..91h4034L}
{Littenberg} T.~B.,  {Cornish} N.~J.,  2015, \mn@doi [\prd]
  {10.1103/PhysRevD.91.084034}, \href
  {https://ui.adsabs.harvard.edu/abs/2015PhRvD..91h4034L} {91, 084034}

\bibitem[\protect\citeauthoryear{{Mandel} \& {Farmer}}{{Mandel} \&
  {Farmer}}{2022}]{2022PhR...955....1M}
{Mandel} I.,  {Farmer} A.,  2022, \mn@doi [\physrep]
  {10.1016/j.physrep.2022.01.003}, \href
  {https://ui.adsabs.harvard.edu/abs/2022PhR...955....1M} {955, 1}

\bibitem[\protect\citeauthoryear{{Pratten}, {Garc{\'\i}a-Quir{\'o}s},
  {Colleoni}  et~al.}{{Pratten} et~al.}{2021}]{2021PhRvD.103j4056P}
{Pratten} G.,  {Garc{\'\i}a-Quir{\'o}s} C.,  {Colleoni} M.,   et~al., 2021,
  \mn@doi [\prd] {10.1103/PhysRevD.103.104056}, \href
  {https://ui.adsabs.harvard.edu/abs/2021PhRvD.103j4056P} {103, 104056}

\bibitem[\protect\citeauthoryear{{Saleem}, {Rana}, {Gayathri}  et~al.}{{Saleem}
  et~al.}{2022}]{2022CQGra..39b5004S}
{Saleem} M.,  {Rana} J.,  {Gayathri} V.,   et~al., 2022, \mn@doi [Classical and
  Quantum Gravity] {10.1088/1361-6382/ac3b99}, \href
  {https://ui.adsabs.harvard.edu/abs/2022CQGra..39b5004S} {39, 025004}

\bibitem[\protect\citeauthoryear{{Samsing} \& {Ilan}}{{Samsing} \&
  {Ilan}}{2019}]{2019MNRAS.482...30S}
{Samsing} J.,  {Ilan} T.,  2019, \mn@doi [\mnras] {10.1093/mnras/sty2249},
  \href {https://ui.adsabs.harvard.edu/abs/2019MNRAS.482...30S} {482, 30}

\bibitem[\protect\citeauthoryear{{Samsing}, {Bartos}, {D'Orazio}
  et~al.}{{Samsing} et~al.}{2022}]{2022Natur.603..237S}
{Samsing} J.,  {Bartos} I.,  {D'Orazio} D.~J.,   et~al., 2022, \mn@doi [\nat]
  {10.1038/s41586-021-04333-1}, \href
  {https://ui.adsabs.harvard.edu/abs/2022Natur.603..237S} {603, 237}

\bibitem[\protect\citeauthoryear{{Speagle}}{{Speagle}}{2020}]{2020MNRAS.493.3132S}
{Speagle} J.~S.,  2020, \mn@doi [\mnras] {10.1093/mnras/staa278}, \href
  {https://ui.adsabs.harvard.edu/abs/2020MNRAS.493.3132S} {493, 3132}

\bibitem[\protect\citeauthoryear{{Varma}, {Biscoveanu}, {Isi}  et~al.}{{Varma}
  et~al.}{2022}]{2022PhRvL.128c1101V}
{Varma} V.,  {Biscoveanu} S.,  {Isi} M.,   et~al., 2022, \mn@doi [\prl]
  {10.1103/PhysRevLett.128.031101}, \href
  {https://ui.adsabs.harvard.edu/abs/2022PhRvL.128c1101V} {128, 031101}

\bibitem[\protect\citeauthoryear{{Veitch}, {Raymond}, {Farr}  et~al.}{{Veitch}
  et~al.}{2015}]{2015PhRvD..91d2003V}
{Veitch} J.,  {Raymond} V.,  {Farr} B.,   et~al., 2015, \mn@doi [\prd]
  {10.1103/PhysRevD.91.042003}, \href
  {https://ui.adsabs.harvard.edu/abs/2015PhRvD..91d2003V} {91, 042003}

\bibitem[\protect\citeauthoryear{{Veske}, {M{\'a}rka}, {Sullivan}
  et~al.}{{Veske} et~al.}{2020}]{2020MNRAS.498L..46V}
{Veske} D.,  {M{\'a}rka} Z.,  {Sullivan} A.~G.,   et~al., 2020, \mn@doi
  [\mnras] {10.1093/mnrasl/slaa123}, \href
  {https://ui.adsabs.harvard.edu/abs/2020MNRAS.498L..46V} {498, L46}

\bibitem[\protect\citeauthoryear{{Veske}, {Sullivan}, {M{\'a}rka}
  et~al.}{{Veske} et~al.}{2021}]{2021ApJ...907L..48V}
{Veske} D.,  {Sullivan} A.~G.,  {M{\'a}rka} Z.,   et~al., 2021, \mn@doi [\apjl]
  {10.3847/2041-8213/abd721}, \href
  {https://ui.adsabs.harvard.edu/abs/2021ApJ...907L..48V} {907, L48}

\bibitem[\protect\citeauthoryear{{Vigna-G{\'o}mez}, {Toonen}, {Ramirez-Ruiz}
  et~al.}{{Vigna-G{\'o}mez} et~al.}{2021}]{2021ApJ...907L..19V}
{Vigna-G{\'o}mez} A.,  {Toonen} S.,  {Ramirez-Ruiz} E.,   et~al., 2021, \mn@doi
  [\apjl] {10.3847/2041-8213/abd5b7}, \href
  {https://ui.adsabs.harvard.edu/abs/2021ApJ...907L..19V} {907, L19}

\bibitem[\protect\citeauthoryear{{Wang}, {Li}, {Vink}, {Fan}, {Tang}, {Qin}  \&
  {Wei}}{{Wang} et~al.}{2022}]{2022ApJ...941L..39W}
{Wang} Y.-Z.,  {Li} Y.-J.,  {Vink} J.~S.,  {Fan} Y.-Z.,  {Tang} S.-P.,  {Qin}
  Y.,   {Wei} D.-M.,  2022, \mn@doi [\apjl] {10.3847/2041-8213/aca89f}, \href
  {https://ui.adsabs.harvard.edu/abs/2022ApJ...941L..39W} {941, L39}

\bibitem[\protect\citeauthoryear{{Xu} \& {Hamilton}}{{Xu} \&
  {Hamilton}}{2023}]{2023PhRvD.107j3049X}
{Xu} Y.,  {Hamilton} E.,  2023, \mn@doi [\prd] {10.1103/PhysRevD.107.103049},
  \href {https://ui.adsabs.harvard.edu/abs/2023PhRvD.107j3049X} {107, 103049}

\bibitem[\protect\citeauthoryear{{Yang}, {Bartos}, {Gayathri}  et~al.}{{Yang}
  et~al.}{2019}]{2019PhRvL.123r1101Y}
{Yang} Y.,  {Bartos} I.,  {Gayathri} V.,   et~al., 2019, \mn@doi [\prl]
  {10.1103/PhysRevLett.123.181101}, \href
  {https://ui.adsabs.harvard.edu/abs/2019PhRvL.123r1101Y} {123, 181101}

\bibitem[\protect\citeauthoryear{{Zackay}, {Dai}, {Venumadhav}, {Roulet}  \&
  {Zaldarriaga}}{{Zackay} et~al.}{2021}]{2021PhRvD.104f3030Z}
{Zackay} B.,  {Dai} L.,  {Venumadhav} T.,  {Roulet} J.,   {Zaldarriaga} M.,
  2021, \mn@doi [\prd] {10.1103/PhysRevD.104.063030}, \href
  {https://ui.adsabs.harvard.edu/abs/2021PhRvD.104f3030Z} {104, 063030}

\makeatother
\end{thebibliography}

\bsp
\label{lastpage}
\end{document}